\documentclass[12pt]{article}

\usepackage{amssymb}
\usepackage{amsthm}
\usepackage{amsmath}

\newtheorem{lemma}{Lemma}[section]
\newtheorem{theorem}{Theorem}[section]
\newtheorem{proposition}{Proposition}[section]

\theoremstyle{definition}
\newtheorem*{remark}{Remark}

\numberwithin{equation}{section}

\newcommand{\esssup}{\operatornamewithlimits{ess\,sup}}

\begin{document}

\title{\textsf{\textbf{The Stochastic Balance Equation for the American Option Value Function and its Gradient}}}

\author{\textsf{\textbf{Malkhaz Shashiashvili}}}

\date{\small School of Mathematics and Computer Science, Kutaisi International University, 33a, Ilia Chavchavadze Ave., Floor 4, Apartment 53, Tbilisi 0179, Georgia \\
E-mail address: \textit{malkhaz.shashiashvili@kiu.edu.ge}}

\maketitle

\begin{abstract}
In the paper we consider the problem of valuation and hedging of American options written on dividend-paying assets whose price dynamics follow the multidimensional diffusion model. We derive a stochastic balance equation for the American option value function and its gradient. We prove that the latter pair is the unique solution of the stochastic balance equation as a result of the uniqueness in the related adapted future-supremum problem.

\vskip+0.5cm


\medskip

\end{abstract}

\section{Introduction}
\label{sec:1}

In this paper we study American options written on dividend-paying assets. We assume that the underlying asset dynamics follow the multidimensional diffusion model. It is well known that the arbitrage-free value of American option can be expressed in terms of the optimal stopping problem (Bensoussan \cite{1}, Karatzas \cite{2}). We consider American options with convex Lipschitz continuous payoff functions $\psi(x)$, $x=(x^1,\dots,x^n)\in[0,\infty)^n$. The class of options with this kind of payoff functions includes index options, spread options, call on max options, put on min options, multiple strike options and others.
We will assume that the time horizon $T$ is finite $T<\infty$, and under the risk-neutral probability measure $P$ the underlying stock prices $S_t=(S_t^1,\dots,S_t^n)$, $0\leq t\leq T$ evolve according to stochastic differential equation
\begin{equation}\label{eq:1.1}
\begin{gathered}
    dS_t^i=S_t^i\big(r-d^{\,i}(t,S_t)\big)\,dt \\
    +S_t^i\Big(\sum_{j=1}^n \sigma_{ij}(t,S_t)\,dW_t^j\Big), \;\; i=1,\dots,n, \;\; 0\leq t\leq T, \\
    \text{with the initial condition} \;\; S_0^i=S^i>0, \;\; i=1,\dots,n.
\end{gathered}
\end{equation}
Here $W_t=(W_t^1,\dots,W_t^n)$, $0\leq t\leq T$, is a standard $n$-dimensional Wiener process on the probability space $(\Omega,\mathcal{F},\mathcal{F}_t,P)$, where $(\mathcal{F}_t)_{0\leq t\leq T}$ is a completion of the natural filtration $(\mathcal{F}_t^W)_{0\leq t\leq T}$ by all $P$-null sets $A\in\mathcal{F}$ with $P(A)=0$.

Further, $r\geq 0$ is the constant rate of interest, $d^{\,i}(t,S)$ is the dividend yield for the stock $S^i$, $i=1,\dots,n$ and $\sigma(t,S)=(\sigma_{ij}(t,S))$, $i,j=1,\dots,n$ is the $n$-dimensional local volatility matrix.

We assume that there exist continuous bounded functions $\widehat{d}^{\,i}(t,x)$, $i=1,\dots,n$, $\widehat{\sigma}(t,x)=(\widehat{\sigma}_{ij}(t,x))$, $i,j=1,\dots,n$, $(t,x)\in[0,T]\times(-\infty,\infty)^n$, such that
\begin{equation}\label{eq:1.2}
\begin{gathered}
    \widehat{d}^{\,i}(t,x)\geq 0, \;\; i=1,\dots,n, \\
    \big|\widehat{d}^{\,i}(t,x)-\widehat{d}^{\,i}(u,y)\big|\leq c\big(|t-u|^{1/2}+|x-y|\big), \;\; i=1,\dots,n, \\
    \big|\widehat{\sigma}_{ij}(t,x)-\widehat{\sigma}_{ij}(u,y)\big|\leq c\big(|t-u|^{1/2}+|x-y|\big), \;\; i,j=1,\dots,n,
\end{gathered}
\end{equation}
with $c\geq 0$, and the matrix $\widehat{a}(t,x)=\widehat{\sigma}(t,x)\widehat{\sigma}^{\top}(t,x)$ is uniformly positive definite:
\begin{equation}\label{eq:1.3}
    \sum_{i,j=1}^n \widehat{a}_{ij}(t,x)y_iy_j\geq\lambda\sum_{i=1}^n y_i^2, \;\; \lambda>0, \;\;\forall\,x,y\in(-\infty,\infty)^n,
\end{equation}
and the latter functions are related with the dividend yield and the local volatility functions by the following change of variable relationship:
for $S=(S^1,\dots,S^n)$ with $S^i>0$, $i=1,\dots,n$, we have
\begin{equation}\label{eq:1.4}
\begin{gathered}
    d^{\,i}(t,S)=\widehat{d}^{\,i}(t,\ln S), \;\; i=1,\dots,n, \\
    \sigma_{ij}(t,S)=\widehat{\sigma}_{ij}(t,\ln S), \;\; i,j=1,\dots,n, \\
    \text{where} \;\; \ln S=(\ln S^1,\dots,\ln S^n).
\end{gathered}
\end{equation}

We will need to introduce several spaces of real valued functions.

$C=C([0,T]\times[0,\infty)^n)$ the space of continuous  real valued functions $f=f(t,S)$, $0\leq t\leq T$, $S=(S^1,\dots,S^n)\in[0,\infty)^n$.

We say that $f\in C$ is Lipschitz continuous in $S\in[0,\infty)^n$, uniformly in $t$, if there exists a constant $L\geq 0$, such that
\begin{equation}\label{eq:1.5}
    \big|f(t,S)-f(t,\widetilde{S})\big|\leq L|S-\widetilde{S}|, \;\; 0\leq t\leq T, \;\; S,\widetilde{S}\in [0,\infty)^n.
\end{equation}

We denote $L^{\infty}((0,T)\times(0,\infty)^n)$ the space of functions $\varphi=\varphi(t,S)$, such that
\begin{equation}\label{eq:1.6}
    |\varphi(t,S)|\leq K \;\;\text{a.e. in $dt\times dS$ for some $K\geq 0$}.
\end{equation}

If $f=f(t,S)\in C([0,T]\times[0,\infty)^n)$ is continuously differentiable with respect to $S$ for each $t$, $0\leq t<T$, then we write
\begin{equation}\label{eq:1.7}
    \nabla f(t,S)=\Big(\frac{\partial f(t,S)}{\partial S^1}\,,\dots,\frac{\partial f(t,S)}{\partial S^n}\Big) \;\;\text{for the gradient of $f=f(t,S)$}.
\end{equation}
$\nabla f\in L^{\infty}$ means that $\frac{\partial f}{\partial S^i}\in L^{\infty}$ for every $i=1,\dots,n$.

We introduce also the parabolic differential operator $Lf(t,S)$ associated with the multidimensional stock prices process $S_t=(S_t^1,\dots,S_t^n)$, $0\leq t\leq T$, satisfying the stochastic differential equation \eqref{eq:1.1}
\begin{align}
    Lf(t,S) & =\frac{1}{2}\sum_{i,j=1}^n a_{ij}(t,S)S^iS^j\,\frac{\partial^2f}{\partial S^i\partial S^j} \nonumber \\
    & +\sum_{i=1}^n \big(r-d^{\,i}(t,S)\big)S^i\,\frac{\partial f}{\partial S^i}+\frac{\partial f}{\partial t}-rf, \label{eq:1.8}
\end{align}
where
$$  a_{ij}(t,S)=\widehat{a}_{ij}(t,\ln S), \;\; i,j=1,\dots,n       $$
and also
$$  a(t,S)=\sigma(t,S)\cdot\sigma^{\top}(t,S).      $$
Consider now the solution of stochastic differential equation \eqref{eq:1.1} $S_u(t,S)=(S_u^1(t,S),\dots,S_u^n(t,S))$, $t\leq u\leq T$, started at time $t$ from the position $S\in (0,\infty)^n$, that is $S_t(t,S)=S$ and define the value function $v(t,S)$, $0\leq t\leq T$, $S\in(0,\infty)^n$, of the following optimal stopping problem
\begin{equation}\label{eq:1.9}
    v(t,S)=\sup_{t\leq\tau\leq T} E^p\big(e^{-r(\tau-t)}\psi(S_{\tau}(t,S))\big),
\end{equation}
where the supremum is taken over all $(\mathcal{F}_u)$, $t\leq u\leq T$ stopping times $\tau$, $t\leq\tau\leq T$.As already mentioned above, it was discovered by Bensoussan \cite{1} and Karatzas \cite{2}, that the arbitrage-free value of American options with the payoff function $\psi(S)$, $S\in(0,\infty)^n$, written on the underlying stock prices $S_t=(S_t^1,\dots,S_t^n)$, $0\leq t\leq T$, coincides with the value function $v(t,S)$ of the optimal stopping problem \eqref{eq:1.9}. For the rigorous mathematical proof of this and some other facts we will refer to the monograph by Pascucci \cite[Chapters~8, 9 and 11]{3}.

For arbitrary function $\varphi(t,S)=(\varphi^1(t,S),\dots,\varphi^n(t,S))\in L^{\infty}((0,T)\times(0,\infty)^n)$ let us introduce the notation
\begin{equation}\label{eq:1.10}
\begin{aligned}
    z^{\varphi}(\theta,S_{\theta}) & =\big(S_{\theta}^1\cdot\varphi^1(\theta,S_{\theta}),\dots,S_{\theta}^n\cdot\varphi^n(\theta,S_{\theta})\big), \;\; 0\leq\theta\leq T, \\
    z^{\nabla v}(\theta,S_{\theta}) & =\Big(S_{\theta}^1\cdot\frac{\partial v}{\partial S^1}\,(\theta,S_{\theta}),\dots,
            S_{\theta}^n\cdot\frac{\partial v}{\partial S^n}\,(\theta,S_{\theta})\Big), \;\; 0\leq\theta\leq T.
\end{aligned}
\end{equation}

Throughout the paper we assume that the payoff function
$$  \psi(S)=\psi(S^1,\dots,S^n), \;\; S=(S^1,\dots,S^n)\in(0,\infty)^n      $$
is the convex Lipschitz continuous function.

\begin{theorem}[The stochastic balance equation]\label{th:1}
Suppose the conditions \eqref{eq:1.2}--\eqref{eq:1.4} are satisfied. Then the pair $(v(t,S),\nabla v(t,S))$ is the unique solution -- pair $(f(t,S),\varphi(t,S))$, where
\begin{gather*}
    f(t,S)\in C([0,T]\times[0,\infty)^n), \quad \varphi(t,S)=(\varphi^1(t,S),\dots,\varphi^n(t,S)), \\
    \varphi^i(t,S)\in L^{\infty}((0,T)\times(0,\infty)^n), \;\; i=1,\dots,n,
\end{gather*}
of the following stochastic balance equation for the American option value function and its gradient
\begin{multline}
    f(t,S_t)=\sup_{t\leq u\leq T} \bigg[e^{-r(u-t)}\psi(S_u) \\
    -\int\limits_t^u e^{-r(\theta-t)}z^{\varphi}(\theta,S_{\theta})\sigma(\theta,S_{\theta})\,dW_{\theta}\bigg], \;\; 0\leq t\leq T, \label{eq:1.11}
\end{multline}
where the multidimensional stock prices process $S_t=(S_t^1,\dots,S_t^n)$, $0\leq t\leq T$, satisfies the stochastic differential equation \eqref{eq:1.1} with the initial condition $S_0=S$, $\forall\,S\in (0,\infty)^n$.
\end{theorem}

\begin{remark}
We would like to note that an inspiration to derive the stochastic balance equation came to us after careful reading of the paper by Davis and Karatzas \cite{4}, in which, in the proof of Theorem 3 the so called future-supremum process appeared for the first time in the literature on optimal stopping problems.
\end{remark}

For the American option value function $v(t,S)$ and its gradient $\nabla v(t,S)$ the stochastic balance equation takes the following form
\begin{multline}
    v(t,S_t)=\sup_{t\leq u\leq T} \bigg[e^{-r(u-t)}\psi(S_u) \\
    -\int\limits_t^u e^{-r(\theta-t)}\Big(S_{\theta}^1\,\frac{\partial v}{\partial S^1}\,(\theta,S_{\theta})\,,\dots,
        S_{\theta}^n\,\frac{\partial v}{\partial S^n}\,(\theta,S_{\theta})\Big)\sigma(\theta,S_{\theta})\,dW_{\theta}\bigg], \;\; 0\leq t\leq T. \label{eq:1.12}
\end{multline}
Notice that equation \eqref{eq:1.12} does not contain partial derivative with respect to time $\frac{\partial v}{\partial t}$\,, nor the second order partial derivatives $\frac{\partial^2v}{\partial S^i\partial S^j}$ with respect to the state argument $S=(S^1,\dots,S^n)$.

The proof of Theorem \ref{th:1} is based essentially on the introduction and analyses of the new problem in stochastic analyses which we call the adapted future-supremum problem.

\section{The adapted future-supremum problem}
\label{sec:2}

Let $(\Omega,\mathcal{F},P)$ be a complete probability space with a filtration $(\mathcal{F}_t)_{0\leq t\leq T}$ satisfying the usual conditions (that is, $\mathcal{F}_0$ contains all $P$-null sets and $\mathcal{F}_t=\mathcal{F}_{t+}$, $0\leq t\leq T$). Consider a real-valued stochastic process $X=(X_t)$, $0\leq t\leq T$, adapted to the given filtration $(\mathcal{F}_t)_{0\leq t\leq T}$ with trajectories $X_t(\omega)$, $0\leq t\leq T$, which are right continuous and possess left limits. We shall assume that the following basic condition is satisfied: there exists a nonnegative uniformly integrable martingale $U=(U_t)$, $0\leq t\leq T$, adapted to the filtration $(\mathcal{F}_t)_{0\leq t\leq T}$, such that
\begin{equation}\label{eq:2.1}
    |X_t|\leq U_t, \;\; 0\leq t\leq T,
\end{equation}
it is well known in this case (see Thompson \cite{6})that there exists a smallest supermartingale $Y=(Y_t)$, $0\leq t\leq T$, greater or equal then $X$, which is called the Snell envelope of $X$ and which possesses the following representation as the value process of the optimal stopping problem
\begin{equation}\label{eq:2.2}
    Y_t=\esssup_{\tau_t} E(X_{\tau_t}\mid\mathcal{F}_t) \;\;\text{for fixed $t$}, \;\; 0\leq t\leq T, \;\; \text{$P$-a.s.},
\end{equation}
where the essential supremum is taken over all $(\mathcal{F}_t)_{0\leq t\leq T}$-stopping times $\tau_t$ with $t\leq \tau_t\leq T$. It is easy to see, that under condition \eqref{eq:2.1}the supermartingale $(Y_t,\mathcal{F}_t)$, $0\leq t\leq T$, is of class $(D)$ and therefore it has a Doob--Meyer decomposition
\begin{equation}\label{eq:2.3}
    Y_t=M_t+B_t, \;\; 0\leq t\leq T,
\end{equation}
where $M=(M_t,\mathcal{F}_t)$, $0\leq t\leq T$, is a uniformly integrable martingale with the initial value $M_0=0$ and $B=(B_t,\mathcal{F}_t)$, $0\leq t\leq T$, is an integrable predictable nonincreasing process and such a decomposition is unique.

For the stochastic process $X=(X_t,\mathcal{F}_t)$, $0\leq t\leq T$, satisfying condition \eqref{eq:2.1} let us introduce the so-called future-supremum process
\begin{equation}\label{2.4}
    \overline{C}_t=\sup_{t\leq s\leq T} X_s, \;\; 0\leq t\leq T.
\end{equation}

It is evident that in general $\overline{C}_t$ is not an $(\mathcal{F}_t)_{0\leq t\leq T}$-adapted process. Indeed, consider the case when the process $X_t$, $0\leq t\leq T$ is increasing. Then we have $\overline{C}_t=X_T$, $0\leq t\leq T$, but $X_T$ is not $(\mathcal{F}_t)_{0\leq t\leq T}$-adapted. Next, let us assume that the process $X_t$, $0\leq t\leq T$, is nonincreasing, then we obtain
$$  \overline{C}_t=X_t, \;\; 0\leq t\leq T,      $$
and hence the process $\overline{C}_t$ is $(\mathcal{F}_t)_{0\leq t\leq T}$-adapted.

Now the question arises: can we adjust the process $X_t$, $0\leq t\leq T$, by a zero mean martingale $M=(M_t,\mathcal{F}_t)$, $0\leq t\leq T$, $M_0=0$, in such a way, that the corresponding future-supremum process
\begin{equation}\label{eq:2.5}
    C_t=\sup_{t\leq s\leq T} (X_s-M_s), \;\; 0\leq t\leq T,
\end{equation}
becomes $(\mathcal{F}_t)_{0\leq t\leq T}$-adapted?

Consider the martingale $M_t=E(X_T\mid\mathcal{F}_t)-E(X_T\mid\mathcal{F}_0)$, $0\leq t\leq T$, $M_0=0$ and let us consider the future-supremum process
$$  C_t=\sup_{t\leq s\leq T} (X_s-M_s)=E(X_T\mid\mathcal{F}_0)+\sup_{t\leq s\leq T} \big(X_s-E(X_T\mid\mathcal{F}_s)\big).      $$

If the initial process $(X_t,\mathcal{F}_t)$, $0\leq t\leq T$, is a submartingale (in particular, increasing process), then it is evident that
$$  X_s-E(X_T\mid\mathcal{F}_s)\leq 0, \;\; 0\leq s\leq T,      $$
while at $s=T$, $X_T-E(X_T\mid\mathcal{F}_T)=0$, hence $C_t=E(X_T\mid\mathcal{F}_0)$ is $(\mathcal{F}_t)_{0\leq t\leq T}$-adapted.

It is a remarkable and intuitively unexpected fact that the following proposition holds true in a quite general framework.

\begin{theorem}\label{th:2}
Let $X=(X_t,\mathcal{F}_t)_{0\leq t\leq T}$ be a real valued stochastic process having right-continuous trajectories $X_t(\omega)$, $0\leq t\leq T$, with left-hand limits, satisfying condition \eqref{eq:2.1}. Then there exists one and only one martingale $(M_t,\mathcal{F}_t)_{0\leq t\leq T}$ with $M_0=0$, such that the future-supremum process,
\begin{equation}\label{eq:2.6}
    C_t=\sup_{t\leq s\leq T} (X_s-M_s), \;\; 0\leq t\leq T, \;\;\text{is $(\mathcal{F}_t)_{0\leq t\leq T}$-predictable}.
\end{equation}
\end{theorem}

We will show at first that the right-continuity of the process $(X-M)$ implies the right-continuity of the nonincreasing process $C$.

\begin{lemma}\label{lem:1}
The nonincreasing future--supremum process $C=(C_t)_{0\leq t\leq T}$ is right continuous, that is
\begin{equation}\label{eq:2.7}
    C_{t+}=C_t, \;\; 0\leq t<T.
\end{equation}
\end{lemma}

\begin{proof}[\textbf{Proof.}]
From the definition of the process $C$ we have
\begin{equation}\label{eq:2.8}
    C_t=\max\Big(\sup_{t\leq s\leq t+\delta} (X_s-M_s),C_{t+\delta}\Big), \;\; 0\leq t<t+\delta\leq T.
\end{equation}
We pass to limit $\delta\downarrow 0$ and get
\begin{equation}\label{eq:2.9}
    C_t=\max(X_t-M_t,C_{t+}), \;\; 0\leq t<T.
\end{equation}
If $C_t>X_t-M_t$, then $C_t=C_{t+}$, hence we should consider the case $C_t=X_t-M_t$. In this case we have
$$  X_t-M_t=C_t\geq C_{t+}\geq C_{t+\delta}\geq X_{t+\delta}-M_{t+\delta}.       $$
After passing to limit $\delta\downarrow 0$ in the latter chain of inequalities, we shall get
$$  X_t-M_t=C_t\geq C_{t+}\geq X_t-M_t,     $$
thus $C_t=C_{t+}$ as required.
\end{proof}

\begin{proof}[\textbf{Proof of Theorem \ref{th:2}.}]
\ \ \

\medskip
\noindent \textit{\textbf{(a) Existence of the desired martingale.}}

Consider the Snell envelope $Y=(Y_t,\mathcal{F}_t)$, $0\leq t\leq T$, of the stochastic process $X=(X_t,\mathcal{F}_t)$, $0\leq t\leq T$. It has the Doob--Meyer decomposition \eqref{eq:2.3}
$$  Y_t=M_t+B_t, \;\; 0\leq t\leq T.        $$
We have
\begin{multline}\label{eq:2.10}
    X_T-M_T\leq\sup_{t\leq s\leq T} (X_s-M_s) \\
    \leq\sup_{t\leq s\leq T} (Y_s-M_s)=\sup_{t\leq s\leq T} B_s=B_t, \;\; 0\leq t\leq T,
\end{multline}
hence the random variable
$$  C_t=\sup_{t\leq s\leq T} (X_s-M_s)      $$
is integrable for each $t$, $0\leq t\leq T$.

From the equality \eqref{eq:2.2} we get
\begin{multline*}
    B_t=Y_t-M_t=\esssup_{\tau_t} E(X_{\tau_t}-M_{\tau_t}\mid\mathcal{F}_t) \\
    \leq E\Big(\sup_{t\leq s\leq T} (X_s-M_s)\mid\mathcal{F}_t\Big)=E(C_t\mid\mathcal{F}_t), \;\; 0\leq t\leq T,
\end{multline*}
and, therefore,
\begin{equation}\label{eq:2.11}
    EB_t\leq EC_t, \;\; 0\leq t\leq T.
\end{equation}
We have from the inequalities \eqref{eq:2.10}, \eqref{eq:2.11} for each $t$, $0\leq t\leq T$, the coincidence
$$  C_t=B_t, \;\; \text{$(P$-a.s.)}     $$
but both nonincreasing processes $C=(C_t)_{0\leq t\leq T}$ and $B=(B_t)_{0\leq t\leq T}$ are right continuous, thus they are indistinguishable
\begin{equation}\label{eq:2.12}
    C_t=B_t \;\;\text{for all $t$}, \;\; 0\leq t\leq T \;\; \text{$(P$-a.s.)}
\end{equation}
and as the filtration $(\mathcal{F}_t)_{0\leq t\leq T}$ satisfies the usual conditions, the future-supremum process $C=(C_t)_{0\leq t\leq T}$ turns out to be $(\mathcal{F}_t)_{0\leq t\leq T}$-predictable.

It is interesting to note that the Snell envelope $Y$ of the process $X$ can be written as
\begin{multline}\label{eq:2.13}
    Y_t=M_t+\sup_{t\leq s\leq T} (X_s-M_s) \\
    =\sup_{t\leq s\leq T} (X_s-(M_s-M_t)) \;\; \text{for all $t$}, \;\; 0\leq t\leq T
\end{multline}
(see Shashiashvili \cite{5}).

\medskip
\noindent \textit{\textbf{(b) The uniqueness of the required martingale.}}

Suppose $(\widehat{M}_t,\mathcal{F}_t)_{0\leq t\leq T}$ with $\widehat{M}_0=0$, is a martingale such that the future-supremum process
\begin{equation}\label{eq:2.14}
    \sup_{t\leq s\leq T} (X_s-\widehat{M}_s) \;\;\text{is $(\mathcal{F}_t)$-predictable}.
\end{equation}

We will show that $\widehat{M}_t=M_t$, $0\leq t\leq T$, where $(M_t,\mathcal{F}_t)_{0\leq t\leq T}$ is the martingale part in the Doob--Meyer decomposition \eqref{eq:2.3} of the Snell envelope $(Y_t,\mathcal{F}_t)_{0\leq t\leq T}$ of the stochastic process $(X_t,\mathcal{F}_t)_{0\leq t\leq T}$.

Denote
\begin{equation}\label{eq:2.15}
    \widehat{C}_t=\sup_{t\leq s\leq T} (X_s-\widehat{M}_s), \;\; \widehat{Y}_t=\widehat{M}_t+\widehat{C}_t, \;\; 0\leq t\leq T.
\end{equation}

Then it is evident that as $(\widehat{C}_t,\mathcal{F}_t)_{0\leq t\leq T}$ is the right continuous nonincreasing predictable process, the stochastic process $(\widehat{Y}_t,\mathcal{F}_t)_{0\leq t\leq T}$ is the right continuous (with left-hand  limits) supermartingale and \eqref{eq:2.15} is its Doob--Meyer decomposition.

We have
\begin{equation}\label{eq:2.16}
    \widehat{C}_t\geq X_t-\widehat{M}_t, \;\; \widehat{Y}_t\geq X_t, \;\; 0\leq t\leq T, \;\; \text{$(P$-a.s.)}
\end{equation}
thus $(\widehat{Y}_t,\mathcal{F}_t)_{0\leq t\leq T}$ is a supermartingale which majorizes $(X_t,\mathcal{F}_t)_{0\leq t\leq T}$.

We have from \eqref{eq:2.15}
\begin{equation}\label{eq:2.17}
    \widehat{C}_{t-\delta}=\max\Big(\sup_{t-\delta\leq s<t} (X_s-\widehat{M}_s),\widehat{C}_t\Big), \;\; 0<t-\delta<t\leq T.
\end{equation}
Tend $\delta\downarrow 0$, we will obtain
\begin{equation}\label{eq:2.18}
    \widehat{C}_{t-}=\max(X_{t-}-\widehat{M}_{t-}\,,\widehat{C}_t), \;\; 0<t\leq T.
\end{equation}

Let us show now the following key property
\begin{equation}\label{eq:2.19}
    \int\limits_0^T I_{(\widehat{Y}_{s-}>X_{s-})}\,d\widehat{C}_s=0.
\end{equation}
Take $s$ such that $\widehat{Y}_{s-}>X_{s-}$. Then $\widehat{C}_{s-}=\widehat{Y}_{s-}-\widehat{M}_{s-}>X_{s-}-\widehat{M}_{s-}$ and from the equality \eqref{eq:2.18} we get $\widehat{C}_{s-}=\widehat{C}_s$, that is, $\Delta\widehat{C}_s=0$, otherwise
\begin{equation}\label{eq:2.20}
    I_{(\widehat{Y}_{s-}>X_{s-})}\cdot\Delta\widehat{C}_s=0.
\end{equation}

Consider the set $(s:\;\widehat{Y}_{s-}-X_{s-}>0)$, we have
\begin{multline}
    (s:\widehat{Y}_{s-}-X_{s-}>0)=\Big(s:\;\widehat{Y}_{s-}-X_{s-}>0,\Delta(X-\widehat{M})_s\neq 0\Big) \\
    \cup\Big(s:\;\widehat{Y}_{s-}-X_{s-}>0,\Delta(X-\widehat{M})_s=0\Big). \label{eq:2.21}
\end{multline}
The first set in \eqref{eq:2.21} is countable, hence
\begin{equation}\label{eq:2.22}
    \int\limits_0^T I_{(\widehat{Y}_{s-}>X_{s-},\Delta(X-\widehat{M})_s\neq 0)}\,d\widehat{C}_s=0
\end{equation}
by the equality \eqref{eq:2.20}.

Consider now the second set of \eqref{eq:2.21} and its arbitrary point $s$:
\begin{equation}\label{eq:2.23}
    \widehat{Y}_{s-}-X_{s-}>0, \;\; \Delta(X-\widehat{M})_s=0.
\end{equation}
We have
$$  \widehat{C}_{s-}=\widehat{Y}_{s-}-\widehat{M}_{s-}>X_{s-}-\widehat{M}_{s-},     $$
hence from \eqref{eq:2.18} and \eqref{eq:2.23} we get
\begin{equation}\label{eq:2.24}
    \widehat{C}_s=\widehat{C}_{s-}>X_s-\widehat{M}_s
\end{equation}
thus there exists $\varepsilon>0$, such that
\begin{equation}\label{eq:2.25}
    \widehat{C}_s>X_s-\widehat{M}_s+\varepsilon.
\end{equation}

As a point $s$ is the continuity point of $(X_s-\widehat{M}_s)$, then $\exists\,\delta>0$, such that
\begin{equation}\label{eq:2.26}
    \Big|(X_s-\widehat{M}_s)-\sup_{s-\delta\leq u\leq s+\delta} (X_u-\widehat{M}_u)\Big|<\frac{\varepsilon}{2}\,.
\end{equation}
Therefore, we can write the inequality
$$  \widehat{C}_s>\sup_{s-\delta\leq u\leq s+\delta} (X_u-\widehat{M}_u)+\frac{\varepsilon}{2}\,.   $$
At the same time from the definition \eqref{eq:2.15} we have
\begin{equation}\label{eq:2.27}
\begin{aligned}
    \widehat{C}_{s-\delta} & =\max\Big(\sup_{s-\delta\leq u<s} (X_u-\widehat{M}_u),\widehat{C}_s\Big), \\
    \widehat{C}_s & =\max\Big(\sup_{s\leq u<s+\delta} (X_u-\widehat{M}_u),\widehat{C}_{s+\delta}\Big),
\end{aligned}
\end{equation}
therefore we get
\begin{equation}\label{eq:2.28}
    \widehat{C}_{s-\delta}=\widehat{C}_s=\widehat{C}_{s+\delta}.
\end{equation}

Thus we have the inclusion
\begin{multline*}
    \Big(s:\;\widehat{Y}_{s-}-X_{s-}>0,\Delta(X-\widehat{M})_s=0\Big) \\
    \subseteq \bigcup_{\begin{subarray}{c} r_1<r_2 \\ \text{$r_1$, $r_2$ -- rationals} \end{subarray}}
                        \Big((r_1,r_2):\;\widehat{C}_{r_2}-\widehat{C}_{r_1}=0\Big)
\end{multline*}
from which we obtain
\begin{equation}\label{eq:2.29}
    \int\limits_0^T I_{(\widehat{Y}_{s-}>X_{s-},\Delta(X-\widehat{M})_s=0)}\,d\widehat{C}_s=0
\end{equation}
Adding equalities \eqref{eq:2.22} and \eqref{eq:2.29} we get
\begin{equation}\label{eq:2.30}
    \int\limits_0^T I_{(\widehat{Y}_{s-}>X_{s-})}\,d\widehat{C}_s=0.
\end{equation}

Consider the stopping times for $\varepsilon>0$
$$  \tau_t^{\varepsilon}=\inf\big(s\geq t:\;X_s\geq\widehat{Y}_s-\varepsilon\big), \;\; 0\leq t\leq T.     $$
If $s$ is such that $t<s\leq\tau_t^{\varepsilon}$, then $X_{s-}\leq \widehat{Y}_{s-}-\varepsilon$, that is, $\widehat{Y}_{s-}\geq X_{s-}+\varepsilon>X_{s-}$, hence
\begin{multline*}
    0\leq E\Big(\widehat{Y}_t-\widehat{Y}_{\tau_t^{\varepsilon}}\mid\mathcal{F}_t\Big)=
        E\bigg(\int\limits_t^T I_{(t<s\leq\tau_t^{\varepsilon})}\,d(-\widehat{C}_s)\mid\mathcal{F}_t\bigg) \\
    \leq E\bigg(\int\limits_t^T I_{(\widehat{Y}_{s-}-X_{s-}>0)}\,d(-\widehat{C}_s)\mid\mathcal{F}_t\bigg)=0,
\end{multline*}
hence
\begin{equation}\label{eq:2.31}
    \widehat{Y}_t=E(\widehat{Y}_{\tau_t^{\varepsilon}}\mid\mathcal{F}_t)\leq E(X_{\tau_t^{\varepsilon}}\mid\mathcal{F}_t)+\varepsilon\leq Y_t+\varepsilon.
\end{equation}
Thus we get $\widehat{Y}_t\leq Y_t$, but $(Y_t,\mathcal{F}_t)_{0\leq t\leq T}$ is the smallest supermartingale that majorizes $(X_t,\mathcal{F}_t)_{0\leq t\leq T}$, therefore we have the equality
$$  \widehat{Y}_t=Y_t, \;\; 0\leq t\leq T,      $$
and from the uniqueness of the Doob--Meyer decomposition we obtain
$$  \widehat{M}_t=M_t, \;\; 0\leq t\leq T,      $$
that is, we have shown the uniqueness of the required martingale    \linebreak    $(\widehat{M}_t,\mathcal{F}_t)_{0\leq t\leq T}$.
\end{proof}

\section{The proof of the Theorem \ref{th:1}}
\label{sec:3}

We shall be based on the obstacle problem for parabolic operators considered in Pascucci \cite{3} in Sections 8.2, 9.4 and 11.3. Parabolic Sobolev spaces $S_{loc}^p((0,T)\times(0,\infty)^n)$ are introduced for any $p\geq 1$ in Section 8.2 and the It\^{o} formula (Theorem 9.47) is proved for functions $f(t,S)$ belonging to the space $S_{loc}^p((0,T)\times(0,\infty)^n)$ for the exponents $p>1+\frac{n+2}{2}$\,.

The obstacle problem is formulated in the following manner:

find a function $f(t,S)$, which belongs to the space $S_{loc}^1((0,T)\times(0,\infty)^n)\cap C([0,T]\times[0,\infty)^n)$ and satisfies the equation
\begin{equation}\label{eq:3.1}
    \max\Big\{Lf(t,S),\psi(S)-f(t,S)\Big\}=0 \;\;\text{a.e. $dt\times dS$ in $(0,T)\times(0,\infty)^n$}
\end{equation}
with the terminal condition
\begin{equation}\label{eq:3.2}
    f(T,s)=\psi(S).
\end{equation}

Here $Lf(t,S)$ is the second order parabolic differential operator \eqref{eq:1.8} and such a function $f(t,S)$ is called a strong solution of the obstacle problem. The basic result Theorem~11.13 about the obstacle problem in Pascucci \cite{3} asserts the existence and the uniqueness of the strong solution of the obstacle problem \eqref{eq:3.1},\,\eqref{eq:3.2} belonging to the parabolic Sobolev space $S_{loc}^p((0,T)\times(0,\infty)^n)$ for any $p\geq 1$. Moreover,
\begin{itemize}
\item[{\rm (1)}] for all $(t,S)\in[0,T]\times(0,\infty)^n$ we have
\begin{equation}\label{eq:3.3}
    f(t,S)=v(t,S),
\end{equation}
where $v(t,S)$ (see \eqref{eq:1.9}) is the value function of the American option with the payoff $\psi(S)$,

\item[{\rm (2)}] the function $f(t,S)$ admits spacial gradient
$$  \nabla f(t,S)=\Big(\frac{\partial f(t,S)}{\partial S^1}\,,\dots,\frac{\partial f(t,S)}{\partial S^n}\Big)   $$
in the classical sense and
\begin{equation}\label{eq:3.4}
    \nabla f(t,S)\in C([0,T)\times(0,\infty)^n)\cap L^{\infty}([0,T)\times(0,\infty)^n).
\end{equation}
\end{itemize}

Let us write the It\^{o} formula for the discounted function $e^{-rt}f(t,S)$ and the $n$-dimensional stock prices diffusion process $S_t=(S_t^1,\dots,S_t^n)$, $0\leq t\leq T$
\begin{align}
    & e^{-rt}f(t,S_t)=f(0,S_0)+\int\limits_0^t e^{-ru}\cdot Lf(u,S_u)\,du \nonumber \\
    &\qquad\qquad\qquad\qquad +\int\limits_0^t e^{z-ru}z^{\nabla f}(u,S_u)\sigma(u,S_u)\,dW_u, \;\; 0\leq t\leq T, \label{eq:3.5} \\
    & e^{-ru}f(u,S_u)-e^{-rt}f(t,S_t)=\int\limits_t^u e^{-r\theta}\cdot Lf(\theta,S_{\theta})\,d\theta \nonumber \\
    &\qquad\qquad\qquad\qquad +\int\limits_t^u e^{-r\theta}z^{\nabla f}(\theta,S_{\theta})\sigma(\theta,S_{\theta})\,dW_{\theta}, \;\; t\leq u\leq T, \label{eq:3.6}
\end{align}
where
$$  z^{\nabla f}(\theta,S_{\theta})=\Big(S_{\theta}^1\,\frac{\partial f}{\partial S^1}\,(\theta,S_{\theta}),\dots,
                    S_{\theta}^n\,\frac{\partial f}{\partial S^n}\,(\theta,S_{\theta})\Big), \;\; 0\leq\theta\leq T.     $$
Denote $M^f=(M_t^f,\mathcal{F}_t)_{0\leq t\leq T}$ the martingale part in the It\"{o} formula \eqref{eq:3.6}
\begin{equation}\label{eq:3.7}
    M_t^f=\int\limits_0^t e^{-r\theta}z^{\nabla f}(\theta,S_{\theta})\sigma(\theta,S_{\theta})\,dW_{\theta}, \;\; 0\leq t\leq T.
\end{equation}
Consider the American option discounted payoff process
\begin{equation}\label{eq:3.8}
    X=(X_t,\mathcal{F}_t)_{0\leq t\leq T}, \;\;\text{where}\;\; X_t=e^{-rt}\psi(S_t),
\end{equation}
and its Snell envelope
\begin{equation}\label{eq:3.9}
    Y=(Y_t,\mathcal{F}_t)_{0\leq t\leq T}.
\end{equation}

\begin{proposition}\label{prop:1}
The Snell envelope $Y$ of the American option discounted payoff process $X$ is given in the following form
\begin{equation}\label{eq:3.10}
    Y_t=e^{-rt}f(t,S_t)=e^{-rt}v(t,S_t), \;\; 0\leq t\leq T.
\end{equation}
\end{proposition}

\begin{proof}[\textbf{Proof.}]
We have from \eqref{eq:3.1}, that $f(t,S)\geq \psi(S)$ and
\begin{equation}\label{eq:3.11}
    Lf(t,S)\leq 0, \quad Lf(t,S)I_{(f(t,S)>\psi(S))}=0 \;\;\text{a.e. $dt\times dS$}.
\end{equation}
As $\nabla f(t,S)$ is a bounded function (see \eqref{eq:3.4}) together with the components of the volatility matrix $\sigma(t,S)$, it is easy to see, that the local martingale \eqref{eq:3.7} is actually square integrable martingale and hence the stochastic process $(e^{-rt}f(t,S_t),\mathcal{F}_t)_{0\leq t\leq T}$ is a supermartingale dominating the process $(e^{-rt}\psi(S_t),\mathcal{F}_t)_{0\leq t\leq T}$. Consider the stopping times
\begin{equation}\label{eq:3.12}
    \tau_t(\omega)=\inf\Big\{u\geq t:\;f(u,S_u(\omega))=\psi(S_u(\omega))\Big\}\wedge T, \;\; 0\leq t\leq T,
\end{equation}
and write the equality \eqref{eq:3.6} at stopping time $\tau_t(\omega)$
\begin{multline}
    e^{-r\tau_t}f(\tau_t,S_{\tau_t})-e^{-rt}f(t,S_t) \\
    =\int\limits_t^{\tau_t} e^{-r\theta}\cdot Lf(\theta,S_{\theta})\,d\theta+(M_{\tau_t}^f-M_t^f), \;\; 0\leq t\leq T. \label{eq:3.13}
\end{multline}

Using the definition of the stopping time $\tau_t$ and taking the conditional expectation in \eqref{eq:3.13} with respect to $\mathcal{F}_t$, we get
\begin{multline}
    E\Big(e^{-r\tau_t}\psi(S_{\tau_t})\mid\mathcal{F}_t\Big)-e^{-rt}f(t,S_t) \\
    =E\bigg(\int\limits_t^{\tau_t} e^{-r\theta}\cdot Lf(\theta,S_{\theta})\,d\theta\mid\mathcal{F}_t\bigg), \;\; 0\leq t\leq T. \label{eq:3.14}
\end{multline}
We have
\begin{multline}
    \bigg|\int\limits_t^{\tau_t} e^{-r\theta}\cdot Lf(\theta,S_{\theta})\,d\theta\bigg|\leq
        \int\limits_t^T I_{(\theta<\tau_t)}|Lf(\theta,S_{\theta})|\,d\theta\bigg| \\
    \leq \int\limits_0^T I_{(f(\theta,S_{\theta})>\psi(S_0))}|Lf(\theta,S_{\theta})|\,d\theta. \label{eq:3.15}
\end{multline}

Let us calculate the expectation of the latter integral
\begin{multline}
    E\int\limits_0^T I_{(f(\theta,S_{\theta})>\psi(S_0))}|Lf(\theta,S_{\theta})|\,d\theta \\
    =\int\limits_0^T\int\limits_{(0,\infty)^n} I_{(f(\theta,S)>\psi(S))}|Lf(\theta,S)|p(0,S_0;\theta,S)\;dS\,d\theta=0 \label{eq:3.16}
\end{multline}
according to the property \eqref{eq:3.11}, where $p(0,S_0;\theta,S)$ is the probability density of the random variable $S_{\theta}=(S_{\theta}^1,\dots,S_{\theta}^n)$, $0<\theta<T$.

Thus we conclude that
\begin{equation}\label{eq:3.17}
    \int\limits_0^T I_{(f(\theta,S_{\theta})>\psi(S_{\theta}))}|Lf(\theta,S_{\theta})|\,d\theta=0 \;\; \text{$(P$-a.s.)}
\end{equation}
and hence from the relations \eqref{eq:3.14}, \eqref{eq:3.15} we get
\begin{equation}\label{eq:3.18}
    e^{-rt}f(t,S_t)=E\Big(e^{-r\tau_t}\psi(S_{\tau_t})\mid\mathcal{F}_t\Big)\leq Y_t, \;\; 0\leq t\leq T,
\end{equation}
where $(Y_t,\mathcal{F}_t)_{0\leq t\leq T}$ is the Snell envelope, that is, the smallest supermartingale dominating the process $(e^{-rt}\psi(S_t),\mathcal{F}_t)_{0\leq t\leq T}$.

Ultimately we have $Y_t=e^{-rt}f(t,S_t)$, $0\leq t\leq T$, hence we have shown the assertion \eqref{eq:3.10} of the Proposition \ref{prop:1}.
\end{proof}

Thus we get the following representation of the Snell envelope $Y_t=e^{-rt}v(t,S_t)$, $0\leq t\leq T$, of the discounted payoff process $X_t=e^{-rt}\psi(S_t)$, $0\leq t\leq T$,
\begin{equation}\label{eq:3.19}
    Y_t=e^{-rt}v(t,S_t)=v(0,S)+\int\limits_0^t e^{-ru}\cdot Lv(u,S_u)\,du+M_t^v, \;\; 0\leq t\leq T,
\end{equation}
where
\begin{equation}\label{eq:3.20}
    M_t^v=\int\limits_0^t e^{-r\theta}\cdot z^{\nabla v}(\theta,S_{\theta})\sigma(\theta,S_{\theta})\,dW_{\theta}, \;\; 0\leq t\leq T.
\end{equation}

\medskip
\noindent \textit{\textbf{Now, we are ready to prove Theorem \ref{th:1}.}}

Consider the Doob--Meyer decomposition \eqref{eq:2.3} of the Snell envelope    \linebreak     $(Y_t,\mathcal{F}_t)_{0\leq t\leq T}$
\begin{equation}\label{eq:3.21}
    Y_t=M_t^v+B_t^v, \quad B_t^v=v(0,S)+\int\limits_0^t e^{-ru}\cdot Lv(u,S_u)\,du, \;\; 0\leq t\leq T.
\end{equation}

We know from \eqref{eq:2.12} that
\begin{equation}\label{eq:3.22}
    B_t^v=\sup_{t\leq u\leq T} (X_u-M_u^v)=\sup_{t\leq u\leq T} \big(e^{-ru}\psi(S_u)-M_u^v\big)
\end{equation}
and hence
\begin{equation}\label{eq:3.23}
    Y_t=\sup_{t\leq u\leq T} \Big(e^{-ru}\psi(S_u)-(M_u^v-M_t^v)\Big), \;\; 0\leq t\leq T
\end{equation}
after multiplying the latter equality by $e^{rt}$ we obtain
\begin{gather}\label{eq:3.24}
\ \hskip-0.5cm v(t,S_t)=\sup_{t\leq u\leq T} \bigg(e^{-r(u-t)}\psi(S_u)-
        \int\limits_t^u e^{-r(\theta-t)}z^{\nabla v}(\theta,S_{\theta})\sigma(\theta,S_{\theta})\,dW_{\theta}\bigg), \\
    \ \hskip+7cm 0\leq t\leq T, \nonumber
\end{gather}
which is the stochastic balance equation \eqref{eq:1.11}.

Suppose now that the pair $(f,t,S),\varphi(t,S))$ satisfies the stochastic balance equation \eqref{eq:1.11}. Multiplying this equation by $e^{-rt}$ we get
\begin{equation}\label{eq:3.25}
    e^{-rt}f(t,S_t)=\sup_{t\leq u\leq T} \big[e^{-ru}\psi(S_u)-(M_u-M_t)\big]=M_t+B_t,
\end{equation}
where
\begin{equation}\label{eq:3.26}
\begin{aligned}
    M_t & =\int\limits_0^t e^{-r\theta}z^{\varphi}(\theta,S_{\theta})\sigma(\theta,S_{\theta})\,dW_{\theta}, \;\; 0\leq t\leq T, \\
    B_t & =\sup_{t\leq u\leq T} \big(e^{-ru}\psi(S_u)-M_u\big), \;\; 0\leq t\leq T.
\end{aligned}
\end{equation}
We see from the equality \eqref{eq:3.25} that the future-supremum process $B_t$, $0\leq t\leq T$, is $(\mathcal{F}_t)_{0\leq t\leq T}$-predictable and hence we can apply the uniqueness statement of Theorem \ref{th:2}, which asserts that the martingale $(M_t,\mathcal{F}_t)_{0\leq t\leq T}$ \eqref{eq:3.26} coincides with the martingale part in the Doob--Meyer decomposition of the Snell envelope $e^{-rt}v(t,S_t)$, $0\leq t\leq T$, that is
\begin{multline}
    M_t=M_t^v, \;\; 0\leq t\leq T, \;\; \text{otherwise} \;\;
        \int\limits_0^t e^{-r\theta}z^{\varphi}(\theta,S_{\theta})\sigma(\theta,S_{\theta})\,dW_{\theta} \\
    =\int\limits_0^t e^{-r\theta}z^{\nabla v}(\theta,S_{\theta})\sigma(\theta,S_{\theta})\,dW_{\theta}, \;\; 0\leq t\leq T. \label{eq:3.27}
\end{multline}

Comparing \eqref{eq:3.23} and \eqref{eq:3.25} we conclude
\begin{equation}\label{eq:3.28}
    f(t,S_t)=v(t,S_t), \;\; 0\leq t\leq T,
\end{equation}
but the functions $f(t,S)$ and $v(t,S)$ are continuous and the random variable $S_t$, $t>0$ has a strictly positive density in $(0,\infty)^n$ (see Bogach\"{e}v, R\"{e}kner, Shaposhnikov \cite[Theorem~3.2]{7}), hence we get the coincidence
\begin{equation}\label{eq:3.29}
    f(t,S)=v(t,S), \;\; (t,S)\in [0,T]\times [0,\infty)^n.
\end{equation}

We have from \eqref{eq:3.27}
\begin{equation}\label{eq:3.30}
    E\bigg(\int\limits_0^T e^{-ru}z^{(\varphi-\nabla v)}(u,S_u)\sigma(u,S_u)\,dW_u\bigg)^2=0,
\end{equation}
that is
\begin{equation}\label{eq:3.31}
    E\int\limits_0^T e^{-2ru}\bigg(\sum_{i,j=1}^n a_{ij}(u,S_u)z_i^{(\varphi-\nabla v)}z_j^{(\varphi-\nabla v)}\bigg)\,du=0,
\end{equation}
where
$$  a(u,S)=\sigma(u,S)\sigma^\top(u,S).     $$
We recall now that the symmetric matrix $a(u,S)$ is uniformly positive definite (see condition \eqref{eq:1.3}) and hence we get from the latter equality
\begin{equation}\label{eq:3.32}
    \int\limits_0^T E\big|z^{(\varphi-\nabla u)}(u,S_u)\big|^2\,du=0,
\end{equation}
which can be written in the following form
\begin{equation}\label{eq:3.33}
    \int\limits_0^T \int_{(0,\infty)^n} \big|z^{(\varphi-\nabla u)}(u,S)\big|^2p(0,S_0;u,S)\;dS\,du=0,
\end{equation}
where the transition probability density $p(0,S_0;u,S)$ is strictly positive (see Bogach\"{e}v, R\"{e}kner, Shaposhnikov \cite[Theorem 3.2]{7}) in $(0,T)\times (0,\infty)^n$. Therefore we conclude
\begin{equation}\label{eq:3.34}
    z^{(\varphi-\nabla v)}(u,S)=0 \;\; \text{a.e. $du\times dS$},
\end{equation}
which gives us the uniqueness assertion of Theorem \ref{th:1}:
\begin{equation}\label{eq:3.35}
\begin{aligned}
    f(t,S) & =v(t,S) , \;\; (t,S)\in [0,T]\times [0,\infty)^n, \\
    \varphi(t,S) & =\nabla v(t,S), \;\; \text{a.e. $dt\times dS$ in $(0,T)\times (0,\infty)^n$}. 
\end{aligned}
\end{equation}

\end{document}